\documentclass[twocolumn,showpacs,preprintnumbers,amsmath,amssymb]{revtex4}
\usepackage{graphicx}
\usepackage{dcolumn}
\usepackage{bm}
\begin{document}

\title{2D to 3D transition in soap films demonstrated by microrheology}

\author{V.~Prasad}
\author{Eric R.~Weeks}
\affiliation{ Department of Physics, Emory University, Atlanta, GA
30322 }

\date{\today}

\begin{abstract}
We follow the diffusive motion of colloidal particles of diameter
$d$ in soap films of varying thickness $h$ with fluorescence
microscopy. Diffusion constants are obtained both from one- and
two-particle microrheological measurements of particle motion in
these films. These diffusion constants are related to the surface
viscosity of the interfaces comprising the soap films, by means of
the Trapeznikov approximation [A. A. Trapeznikov, \emph{PICSA}
(1957)] and Saffman's equation for diffusion in a 2D fluid.
Unphysical values of the surface viscosity are found for thick soap
films ($h/d > 7 \pm 3$), indicating a transition from 2D to 3D
behavior.
\end{abstract}

 \maketitle

A soap film is a thin layer of fluid stablized by two surfactant
layers that buffer it from air phases above and below. In the early
18th century, Sir Isaac Newton measured the thickness of the fluid
layer to $\sim 10$ nm precision \cite{Newton}. In fact, because of
the similarity of thin ``Newton black films'' to planar lipid
bilayers, soap films have been proposed as models for cell membranes
\cite{tien1}. The analogy with membranes extends to considering a
thin soap film as a 2D fluid \cite{Cheung1}. This has motivated the
use of soap films to study turbulence in 2D
\cite{goldburg1,swinney1}, as well as informing the physics of
drainage in foams \cite{stone2}. However, soap films have a nonzero
thickness, and presumably under some conditions the model of the
film as a 2D fluid is inappropriate.

A previous study by Cheung \emph{et al.} \cite{Cheung1} quantified
the hydrodynamics of a single soap film for the special case where
the diameter $d$ of embedded tracer particles was assumed to be the
same as the thickness $h$ of the film.  They observed the relative
diffusive motion of pairs of these particles, and found that this
relative diffusion depended logarithmically on the separation
between the particles, indicating 2D fluid-like behavior. Clearly,
for thicker films where $h\gg d$, 3D behavior must be recovered.
This transition from 2D to 3D behavior has not been demonstrated in
any study to date.

In this Letter, we use the thermal motion of embedded polystyrene
particles to study soap films of varying thickness $h$, to clarify
which aspects of particle motion arise from 2D hydrodynamics and
which from 3D hydrodynamics. For small particles in thick films
($h/d>7$), the measured particle diffusivity is similar to free 3D
diffusion in the fluid comprising the film.  For thin films
($h/d<7$), particles diffuse noticeably faster, suggesting that
particle drag is more due to 2D hydrodynamics with an effective 2D
viscosity. Measurements of the correlated motion of pairs of
particles show that {\it all} soap films have 2D-like long-range
correlations. The classic Trapeznikov approximation
\cite{Trapeznikov1} connects the 2D and 3D properties of the film by
modeling the soap film as a 2D interface with an effective surface
viscosity $\eta_{s,\text{eff}}$ given by
\begin{equation}
\eta_{s\text{,eff}}=\eta_{\text{bulk}}h+2\eta_{\text{int}}
\end{equation}
in terms of the 3D viscosity $\eta_{\text{bulk}}$ of the fluid in
the film, and the 2D surface viscosity $\eta_{\text{int}}$ of the
surfactant layers; see Fig.~1.  Our results show that Eq.~(1) and
Fig.~1 are valid for thin films but not for thick films where 3D
hydrodynamics becomes important.  These observations lead us to
conclude that a transition from 2D fluid to 3D bulk behavior occurs
at around $h/d \approx 7 \pm 3$, the first experimental
demonstration of such a transition.

\begin{figure} [htbp]
\includegraphics[scale=0.45]{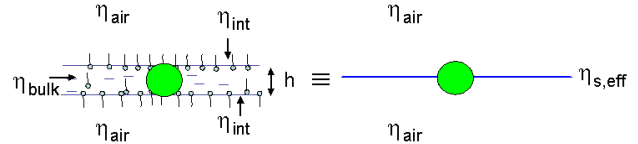}
\caption{\label{fig:0} Schematic of the Trapeznikov approximation,
where the entire soap film is approximated as a single interface in
contact with bulk air phases.}
\end{figure}

We use mixtures of water, glycerol and the commercially available
dishwashing detergent Dawn to prepare our soap films. This
particular brand was chosen as it has proved quite successful at
making long-lasting soap films \cite{Rutgers1}; further, the
preponderance of its use in the literature allows us to compare our
results with previous work \cite{Cheung1,stone2,rutgers2}. The
concentration of Dawn is kept fixed at 2$\%$ by weight in our soap
solutions to try to maintain a constant interfacial viscosity
$\eta_\text{{int}}$ for all films.  The fluid viscosity
$\eta_\text{{bulk}}$ in our films is controlled by changing the
ratio of water and glycerol in the soap solutions.  Fluorescent
polystyrene spheres (Molecular Probes, carboxyl modified, $d=210$ or
$500$ nm) are added to these mixtures to act as tracer particles.

Stable soap films are created by dipping and drawing out a circular
stainless steel frame of thickness 1 mm from the soap solutions. The
steel frame is enclosed in a chamber that minimizes convective
effects in the soap film while maintaining its relative humidity. We
then image the particles by fluorescent microscopy; soap films
containing 500 nm particles are imaged with a 20$\times$ objective
(numerical aperture = 0.4, 465 nm/pixel) while those with 210 nm
particles are imaged with a 40$\times$ objective (numerical aperture
= 0.55, 233 nm/pixel). For each sample, short movies of duration
$\sim$30~s are recorded with a CCD camera that has a 640 $\times$
486 pixel resolution, at a frame rate of 30 Hz. After each movie, we
transfer the film to a spectrophotometer and its thickness $h$ is
determined from the transmitted intensity \cite{huibers1}. The
movies are then analyzed by particle tracking to obtain the
positions of the tracers. From the particle positions, we determine
their displacements by the relation $\Delta
\vec{r}(t,\tau)=\vec{r}(t+\tau)-\vec{r}(t)$, where $t$ is the
absolute time and $\tau$ is the lag time. Finally, any global motion
is subtracted from these displacements to eliminate the remnant
effects of convective drift caused by the air phases that contact
the soap film.

\begin{figure} [htbp]
\includegraphics[scale=0.36]{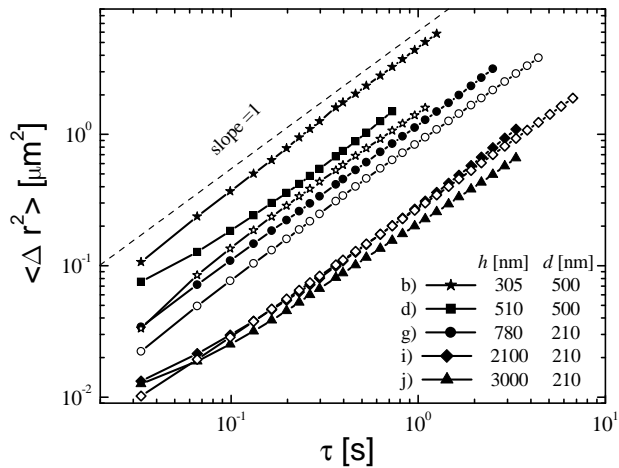}
\caption{\label{fig:1} The solid symbols are the mean square
displacements for five soap films with increasing $h/d$ ratios. See
Table~I for the viscosities $\eta_\text{{bulk}}$ that correspond
to these films. The open symbols represent data from the bulk
(3D) solutions used to make the soap films for four cases (b, g,
i/j). The dashed line with a slope of 1 is a guide to the eye.}
\end{figure}

To quantify particle motion, we use the vector displacements of the
particles to calculate the mean square displacements (MSDs)
$\langle\Delta r^{2}\rangle$ as a function of the lag time $\tau$.
Figure~2 shows measurements of $\langle\Delta r^{2}\rangle$ for five
different soap films, where $h/d$ ranges from 0.6 to 15. The
viscosity $\eta_\text{{bulk}}$ of the fluid layer of these soap
films is given in Table~I.  At long lag times, all the MSDs are
linear with respect to $\tau$, irrespective of $h/d$ and
$\eta_\text{{bulk}}$. This is expected, as both the fluid layer and
the interfaces comprising the soap film are viscous. We extract a
one-particle self-diffusion coefficient, $D_{s,1p}$, from the
measurements according to the equation $\langle\Delta
r^{2}\rangle=4\tau D_{s,1p}$.  Further, from Table~I and Fig.~2 it
is clear that increasing both $\eta_\text{{bulk}}$ and $h/d$ tends
to slow the diffusion of the particles.  Films made from more
viscous bulk fluids tend to be thicker (see Table~I), and so these
single-particle measurements do not clearly distinguish the
influence of $\eta_\text{{bulk}}$ and $h/d$, although it is obvious
that increasing $\eta_\text{{bulk}}$ should slow diffusion, and
plausible that increasing $h/d$ might also slow diffusion. This
latter effect is suggested by comparing films i and j in Fig.~2,
which have the same $\eta_\text{{bulk}}$; the motion is slower for
the thicker film j (upward solid triangles). A further suggestion
that thicker films result in slower diffusion comes from comparing
the motion within the films (solid symbols in Fig.~2) with motion in
the 3D fluid solutions the films are made from (open symbols in
Fig.~2).  For a thin film ($h/d = 0.6$, stars) the particle motion
is much faster in the soap films than in the corresponding 3D
solution. For the thickest films we study ($h/d \approx 10-15$,
solid diamonds and triangles) the motion in the soap film is
comparable to the motion in the corresponding 3D solution (open
diamonds).

\begin{table}
\caption{\label{tab:table1} Parameters for all the soap films
described in this paper. $\eta_\text{{bulk}}$ (determined from
diffusivity measurements in bulk solutions) has an error of $\pm
5\%$, and values of $h$ and $d$ are certain to within $\pm 2\%$. The
uncertainities in $\eta_{\text{int}}$, derived from Eq.~1 and 2, are
given in the brackets.}
\begin{ruledtabular}
\begin{tabular}{lcccc}
$\eta_{\text{bulk}}$&$h$&$d$&$\eta_{\text{int}}(1p)$&$\eta_{\text{int}} (2p)$\\
(mPa$\cdot$s)&(nm)&(nm)&(nPa$\cdot$m$\cdot$s)&(nPa$\cdot$m$\cdot$s)\\
\hline
 a. 2.0 \cite{Cheung1} &400 &400&0.20 ($\pm$ 0.03)&0.47 ($\pm$ 0.06)\\
 b. 2.3&305&500&0.63 ($\pm$ 0.06)&1.02 ($\pm$ 0.10)\\
 c. 3.0&640&500&0.49 ($\pm$ 0.09)&0.62 ($\pm$ 0.12)\\
 d. 6.0&510&500&0.89 ($\pm$ 0.2)&0.84 ($\pm$ 0.2)\\
 e. 10.0&1340&500&0.34 ($\pm$ 0.5)&2.26 ($\pm$ 0.7)\\
 f. 25.0&1100&500&-0.30 ($\pm$ 0.9)&4.35 ($\pm$ 1.5)\\
 g. 10.0&780&210&0.12 ($\pm$ 0.26)&1.64 ($\pm$ 0.4)\\
 h. 25.0&2184&210&-8.92 ($\pm$ 1.3)&27.2 ($\pm$ 4)\\
 i. 30.0&2100&210&-10.6 ($\pm$ 1.5)&25.0 ($\pm$ 4)\\
 j. 30.0&3000&210&-15.5 ($\pm$ 2.1)&65.2 ($\pm$ 8)\\
\end{tabular}
\end{ruledtabular}
\end{table}

To further understand the hydrodynamics and how particle motion
compares in thick and thin films, we use the correlated
motions of particles \cite{crocker1,mason2,prasad1} to
probe flow fields in these soap films.  Briefly, we look
at the product of particle displacements $D_{rr}(R,\tau)
= \langle\Delta r_{r}^{i}(t,\tau)\Delta r_{r}^{j}(t,\tau)
\delta(R-R^{ij}(t))\rangle_{i\neq j,t}$, where \emph{i},
\emph{j} are particle indices, the subscripts $r$ represent
motion parallel to the line joining the centers of particles,
and $R^{ij}$ is the separation between particles $i$ and $j$.
Similar to \cite{prasad1}, we observe $D_{rr} \sim \tau$, which
enables the estimation of a $\tau$-independent quantity $\langle
D_{rr}/\tau\rangle_{\tau}$, depending only on $R$ and having units
of a diffusion constant.

\begin{figure}[htbp]
\includegraphics[scale=0.36]{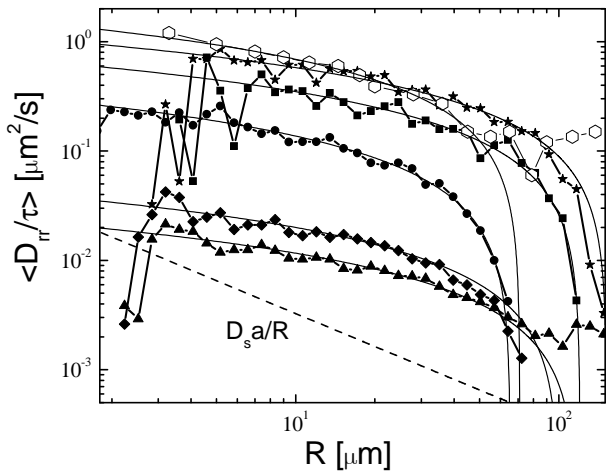}
\caption{\label{fig:2} Two-particle longitudinal correlation
function $\langle D_{rr}/\tau\rangle$ for the five soap films
described in Fig.~2, symbols being the same. An additional data set
from \cite{Cheung1} has been included (open hexagons). Solid lines
are empirical fits to the data of the form $A \text{ ln } R + B$.
The dashed line is the form of the correlation function expected for
a 3D fluid with $\eta_{\text{bulk}}=6.5$ mPa$\cdot$s (squares).}
\end{figure}

In Fig.~3 we show $\langle D_{rr}/\tau\rangle$ as a function of the
separation $R$ for the five soap films described in Fig.~2, with an
additional data set from Ref.~\cite{Cheung1} included. The motion of
a tracer particle creates a flow field in the soap film that affects
the motion of other particles, and the correlation function
indicates the spatial extent of this flow field.  The dashed line in
Fig.~3 represents the form of the correlation function in a 3D fluid
\cite{crocker1}; therefore it is clear that the motion is correlated
over larger distances in soap films than in 3D. This long-ranged
behavior is characteristic of 2D fluids \cite{levine1,stone1}.
Further, similar to the trend seen in the MSDs, increasing $h/d$ and
$\eta_\text{{bulk}}$ lowers the value of the correlation function
$\langle D_{rr}/\tau\rangle$ for the same separation $R$. As
$\langle D_{rr}/\tau\rangle$ dimensionally represents a diffusion
constant, slower diffusion decreases its magnitude. Finally, the
correlation functions for all six soap films are similar in shape.
This is evident from the form of the function, $A \text{ ln } R + B$
that has been used to empirically fit all the six curves.

The data set from Cheung \emph{et al}.~\cite{Cheung1} (open
hexagons) requires explanation. In their paper, the data was
presented as a two-particle MSD, $\langle\Delta R^{2}\rangle =
\langle\{[r_j(t+\tau)-r_i(t+\tau)]-[r_j(t)-r_i(t)]\}^2\rangle$,
which measures the relative diffusion between particles $i$ and $j$.
This was done for a fixed $\tau$ (=1/30 s), and the resulting
relative diffusion decomposed into two components $\langle\Delta
R^2\rangle=\langle \Delta R_{\parallel}^{2}\rangle+\langle \Delta
R_{\perp}^{2}\rangle$, representing displacements parallel and
perpendicular to the lines joining the centers of the particles. It
is then straightforward to show that $\langle D_{rr}/\tau\rangle =
(2\langle\Delta r_r^{2}\rangle-\langle \Delta
R_{\parallel}^{2}\rangle)/2\tau \approx (\langle\Delta
r^{2}\rangle-\langle \Delta R_{\parallel}^{2}\rangle)/2\tau$ (see
\cite{calculation}). We plot the data in this form in Fig.~3,
evaluated at $\tau$ = 1/30~s.

The data shown in Fig.~3 can be used to extract a single-particle
self diffusion constant, $D_{s,2p}$, but now measured from
two-particle correlations. This is done by extrapolating the
correlation functions $\langle D_{rr}/\tau\rangle$ to $R=d/2$. The
reason for this choice is that the single particle diffusion
constant must be recovered from the two-particle measurement when
extrapolated to the particle radius \cite{crocker1}. We then deduce
that $D_{rr}(R=d/2,\tau) =\langle\Delta
r_{r}^2\rangle\approx2D_{s,2p}\tau$ \cite{calculation} and use the
fitting functions shown in Fig.~3 to determine the value of
$D_{s,2p}$ for each soap film. However, this extrapolation process
has limitations: nearby particles at interfaces can have strong
interactions, either electrostatic or through capillary forces. Our
particle concentration was chosen to have spheres be no closer than
$R \sim 5d$, to avoid such effects. The excellent agreement between
the fitting function and all our data gives us confidence, however,
that the extrapolation is valid even when $R< 5d$.

\begin{figure} [htbp]
\includegraphics[scale=0.5]{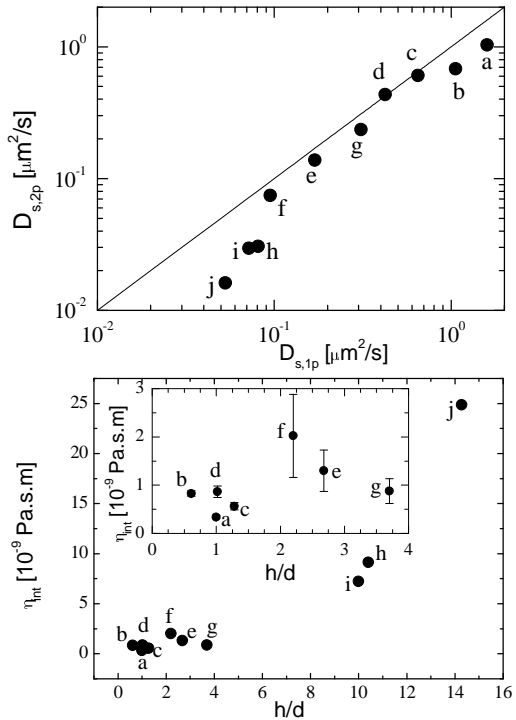}
\caption{\label{fig:4} (a) Two-particle diffusion constant
$D_{s,2p}$ plotted against the one-particle $D_{s,1p}$ for the six
soap films described in Figs.~2 and 3. Four additional data sets
have been included, details of which are given in Table~I. The
straight line indicates equality between the diffusion constants.
Errors in $D_{s,1p}$ and $D_{s,2p}$ are $\pm 5\%$, similar to that
in $\eta_\text{{bulk}}$, and are smaller than the size of the
symbols in the figure. (b) Interfacial viscosity $\eta_\text{{int}}$
as a function of $h/d$. Inset: Magnified view of $\eta_\text{{int}}$
for $h/d < 4$, with error bars included.}
\end{figure}

Insight into the transition from 2D to 3D behavior can be gained by
plotting the diffusion constants obtained from the two methods
against each other, as shown in Fig.~4(a). The two diffusion
constants agree with each other for thin soap films while for
thicker films, the deviation from equality lies beyond experimental
error (points h, i, and j). These results can be interpreted using
the Trapeznikov approximation described in Fig.~1, which models the
soap film as a 2D interface with an effective surface viscosity
$\eta_{\text{int}}$ given by Eq.~(1).  Based on this approximation,
our system reduces to that of a particle diffusing at an interface
in contact with bulk air phases. The diffusion of a disk or sphere
\cite{stone3} at such an interface has been described by Saffman
\cite{saffman1} and is given by
\begin{equation}
D_{s} =
\frac{k_{B}T}{4\pi\eta_{s\text{,eff}}}[\text{ln}
(\frac{2\eta_{s\text{,eff}}}{\eta_{\text{air}} d})-\gamma_{E}]
\end{equation}
where $\eta_\text{{air}}$ is the viscosity of the bulk air
phases and $\gamma_{E}$ is Euler's constant. Equation~(2)
holds if $2\eta_{s,\text{eff}} \gg \eta_{\text{air}}d$, which
is true for our data due to the low value of the air viscosity
($\eta_{\text{air}}=0.017$ mPa$\cdot$s).

We now attempt to determine the interfacial viscosity of the
surfactant layers by using Saffman's equation [Eq.~(2)] to convert
measurements of $D_{s,1p}$ and $D_{s,2p}$ into
$\eta_{s\text{,eff}}$, and then using the Trapeznikov approximation
[Eq.~(1)] to determine
$\eta_\text{{int}}=1/2(\eta_{s,\text{eff}}-\eta_\text{{bulk}}h)$.
For all films, we average $D_{s,1p}$ and $D_{s,2p}$ to determine
$\eta_\text{{int}}$, which is plotted in Fig.~4(b) as a function of
$h/d$.  For the thin films the interfacial viscosity shows roughly
constant behavior while for thick films the variation in
$\eta_\text{{int}}$ is quite pronounced beyond the experimental
uncertainty in the measurements (see Table I). The inset to Fig.~4
(b) shows a magnified view of the interfacial viscosity for $h/d <
4$, where it is clear that $\eta_\text{{int}}$ is nearly constant
with an average value of $0.97 \pm 0.55$ nPa$\cdot$s$\cdot$m. This
is expected because the same concentration of Dawn surfactant has
been used in all these soap films.

For thick films, the one-particle measurements $D_{s,1p}$ give large
\emph{negative} values of $\eta_\text{{int}}$ (refer Table I)
implying that the single particle diffusivities are significantly
faster than that predicted by Eqs.~(1,2).  From Fig.~2, it is clear
that the 3D Stokes-Einstein equation for diffusion,
$D_{s,1p}=k_{B}T/(3\pi\eta_{\text{bulk}}d)$, is sufficient to
explain the motion of the probe particles in thicker soap films,
without the need to invoke Saffman's equation. This makes sense, as
in the limit of a 3D system ($h \rightarrow \infty$), Eq.~(1)
predicts an infinite surface viscosity, which has no physical
meaning.  The apparent negative values of $\eta_\text{{int}}$ for
$h/d > 7$ indicate that the 3D limit is already evident for films of
this thickness. In contrast with the one-particle measurements, the
two-particle measurements in thick films give large {\it positive}
values of $\eta_\text{{int}}$ (see Table I), again contradictory to
the low values determined in thin films.  An alternate way to state
this is that the Trapeznikov approximation predicts an effective
surface viscosity that is too small, if we use $\eta_\text{{int}}$
based on the thin film measurements.

This leaves us with a puzzle; even for these thick films the
two-particle correlation functions are long-ranged indicating that
the soap films behave like a 2D fluid. In fact, the behavior of the
correlation functions as a function of $R$ for all soap films can be
explained by considering the following. Locally, the films likely
behave as a 3D fluid \cite{diamant}. We hypothesize that the
correlation functions in the thick films would then decay as $1/R$
at very short separations ($d/2<R<h$) but more slowly at larger
separations ($R>h$) . The extrapolation of $D_{rr}$ to $R=d/2$ thus
underpredicts $D_{s,2p}$, explaining the overestimation of
$\eta_\text{{int}}$ for the thick films. At intermediate
separations, because of conservation of fluid momentum, \emph{all}
soap films behave as 2D fluids \cite{diamant}. Therefore, the
correlation functions decay in a logarithmic fashion for those
separations, as seen by the form of the fitting functions in Fig.~3.
However, the logarithmic divergence of the correlation function is
cut off at a length scale where stresses in the air phase from
motion of the the tracers become important. This length scale,
related to $\eta_{s,\text{eff}}/\eta_{\text{air}}$, is the
separation at which the correlation functions begin to decay more
rapidly, indicating a final crossover to 3D fluid like behavior.

Our work describes a transition from 2D to 3D behavior when the
thickness of soap films are changed with respect to particle size,
at a ratio of $h/d \approx 7 \pm 3$. This demonstrates that the
hydrodynamics of thin films depend on the ratio of the dimensions of
the film and the probe particle, rather than intrinsically on the
film itself. The particular value of $h/d \approx 7 \pm 3$ is an
empirical determination of when 3D shear gradients in the fluid
layer dominate dissipation of stress from the motion of the probe,
in comparison to the air phase. For thin films showing 2D behavior,
the air phase is crucial for stress dissipation, as demonstrated by
the presence of $\eta_\text{{air}}$ in the Saffman equation
[Eq.~(2)]. In all our films, the interface is relatively
\emph{mobile}, that is, $\eta_\text{{int}}$ is small when compared
to the contribution from the fluid layer $\eta_\text{{bulk}}h$.
Changing the nature of the interface, such as making the interface
more rigid ($\eta_\text{{int}} \gg \eta_\text{{bulk}}h$, for
instance) will likely change where the transition occurs.


\begin{thebibliography}{19}
\bibitem[{cal()}]{calculation}
\bibinfo{note}{In this work, we focus on $D_{rr}$, the correlated motion of two
  particles along the direction of the line separating them. An additional
  component $D_{\theta\theta}$ can be computed, where $\theta$ indicates the
  direction of motion perpendicular to this line. Thus the true displacement of
  a single particle of diameter $d$ should be
  $D_{rr}(R=d/2,\tau)+D_{\theta\theta}(R=d/2,\tau) = \langle \Delta r_r^2
  \rangle + \langle \Delta r_{\theta}^2 \rangle = \langle \Delta r^2 \rangle =
  4 D_{s,2p}\tau$. In 2D, $\langle \Delta r_r^2 \rangle \approx \langle \Delta
  r_{\theta}^2\rangle$ and thus $D_{rr}(R=d/2,\tau) \approx 2 D_{s,2p}\tau$.}
\end{thebibliography}
\end{document}